# Illustrating Dynamical Symmetries in Classical Mechanics:
# The Laplace-Runge-Lenz Vector Revisited

Ross C. O'Connell and Kannan Jagannathan
Physics Department, Amherst College
Amherst, MA 01002-5000

Abstract

The inverse square force law admits a conserved vector that lies in the plane of motion. This vector has been associated with the names of Laplace, Runge, and Lenz, among others. Many workers have explored aspects of the symmetry and degeneracy associated with this vector and with analogous dynamical symmetries. We define a conserved dynamical variable $\alpha$ that characterizes the orientation of the orbit in two-dimensional configuration space for the Kepler problem and an analogous variable $\beta$ for the isotropic harmonic oscillator. This orbit orientation variable is canonically conjugate to the angular momentum component normal to the plane of motion. We explore the canonical one-parameter group of transformations generated by $\alpha$ ($\beta$). Because we have an obvious pair of conserved canonically conjugate variables, it is desirable to use them as a coordinate-mometum pair. In terms of these phase space coordinates, the form of the Hamiltonian is nearly trivial because neither member of the pair can occur explicitly in the Hamiltonian. From these considerations we gain a simple picture of dynamics in phase space. The procedure we use is in the spirit of the Hamilton-Jacobi method.

## I. The Kepler Problem

There are hundreds of papers[1-10] on the power, applications, and extensions of the Laplace-Runge-Lenz vector. The paper by Mukunda[11] comes closest to what we discuss below, but our aim is primarily pedagogical.

Students in intermediate mechanics courses are usually taught that there is a deep connection between symmetries, degeneracy, and conservation laws. A student impressed by this insight might wonder what symmetry, if any, is connected with the Laplace-Runge-Lenz vector. We have not found an answer in the literature that is accessible to advanced undergraduates. To be sure, there are discussions of the dynamical O(4) and SU(3) symmetries of the Kepler and harmonic oscillator problems respectively. For instance, O(4) has six generators, of which three are the components of angular momentum and generate ordinary rotations, and three are related to the components of the Runge-Lenz vector. However, students are not likely to have the background to appreciate the insights provided by this approach.

Although the symmetries of space and time (the kinematic symmetries) imply important conservation laws such as those of energy, momentum, and angular momentum, there are other conserved quantities that follow from symmetries in phase space. Dynamical symmetries are precisely the extra symmetries realized in phase space, and the Hamiltonian formulation of dynamics is most conducive for their investigation. There are excellent general textbooks[12, 13] that offer ways of exploring Hamiltonian flows in phase space and their associated symmetries, but these approaches also may not be accessible at the intermediate mechanics level. In this paper, we take advantage of the fact that students are familiar with the ordinary dynamics of the Kepler problem and the isotropic harmonic oscillator to motivate the consideration of symmetries in the simple phase spaces of these systems. Although we cite the literature for more advanced or general treatments, we hope that the following discussion will be self-contained and introduce the reader to the central and powerful idea of dynamics.

This brief introduction will serve mainly to explain our notation. Let a particle of mass m be subject to an inverse square central force $\frac{-k}{r^2}$; if k > 0, the force is attractive. and if k < 0, it is repulsive. It is not difficult to verify that

$$\vec{A} = \frac{\vec{p} \times \vec{L}}{mk} - \hat{r}, \tag{1}$$

is a constant of the motion, where $\vec{p}, \vec{L}, \hat{r}$ are the momentum, angular momentum, and radial unit vectors, respectively. Moreover, $\vec{A}$ is in the plane of the motion, and its magnitude is the eccentricity of the conic section that is the orbit, and hence $|\vec{A}|$ is determined by the energy and the angular momentum. Thus, given the conservation of energy and angular momentum, the conservation of $\vec{A}$ amounts to the additional information that its *direction* in the plane of motion is a constant of the motion and points from the center of force to the pericenter of the orbit in the attractive case.

The conservation of the angular momentum vector restricts the motion to a plane, the x-y plane, and hence there is no loss if we consider the two-dimensional problem from the start. In this case, we may conveniently regard the conservation of angular momentum to be simply the conservation of what would have been its z-component in the three-dimensional problem,

$$L(x, y, p_x, p_y) = x p_y - y p_x \tag{2}$$

Depending on the sense of motion, L can be positive or negative. The Laplace-Runge-Lenz vector now has two components:

$$A_x = \frac{Lp_y}{mk} - \frac{x}{\sqrt{x^2+y^2}},$$ (3)

and

$$A_y = -\frac{Lp_x}{mk} - \frac{y}{\sqrt{x^2+y^2}}.$$ (4)

Because the components are each constants of the motion, so is the angle between the vector and the x-axis, defined by

$$\alpha(x, y, p_x, p_y) = \tan^{-1}\frac{A_y}{A_x}.$$ (5)

In an obvious manner, α specifies the orientation of the orbit. We have explicitly indicated the four phase space variables as the arguments of α to emphasize that α is a dynamical variable and hence is a function of the state of the system just as L is.

An explicit, though rather tedious calculation establishes the following Poisson bracket:

$$\{\alpha, L\} = 1.$$ (6)

The Poisson brackets with the Hamiltonian can be established either through an explicit calculation or by noting that both α and L are conserved,

$$\{\alpha, H\} = \{L, H\} = 0.$$ (7)

Thus not only can we make a canonical transformation to a new set of coordinates in phase space with α and L as a pair of canonically conjugate variables, but if we do, both members of this pair are conserved. This fact is what is special about the variable α. A number of angle variables, such as θ = tan$^{-1}$(y/x), have the right Poisson bracket with L to constitute potentially a canonically conjugate pair, but in general, they would not also be conserved.

Suppose in addition to L and α, Q and P are a second pair of conjugate variables after the canonical transformation. Then the Hamiltonian can be written as a function of the new phase space variables (Q, P, α, L). On the one hand, the conservation of α and L implies that the Hamiltonian written in the new coordinates should not explicitly contain either of those variables:

$$0 = \dot{\alpha} = \frac{\partial H}{\partial L}, \text{ and } 0 = \dot{L} = -\frac{\partial H}{\partial \alpha}. \tag{8}$$

On the other hand, {L , Q} = {α, Q} = {L , P} = {α, P} = 0. An obvious choice that meets all these criteria is to choose H as the new P. Then in the new coordinates,

$$H = P,$$
$$\dot{Q} = \frac{\partial H}{\partial P} = 1. \tag{9}$$

Thus

$$Q(t) = Q_0 + t. \tag{10}$$

Q is the variable along each curve of motion parametrized by t except for an arbitrary additive constant.

The general scheme outlined here is well known in classical mechanics: for a system with n degrees of freedom and a time independent Lagrangian, we can find (2n-1) time-independent constants of motion as the coordinates of phase space. The remaining coordinate will turn out to be proportional to time. One typically sees this general development in the context of the Hamilton-Jacobi formulation. We have discussed an example of the general result in a non-trivial context that can be usefully introduced in intermediate mechanics courses.

Next we try to understand the role of α as the infinitesimal generator of a one-parameter subgroup (of canonical transformations) in phase space. The general principles are well known: Just as L generates rotations in the x-y plane (to be illustrated explicitly below), and just as the Hamiltonian generates time evolution, then for any dynamical variable $G(x, y, p_x, p_y)$, we can find the one-parameter group of transformations generated by it in phase space (at least locally).

There is an informal way to understand the nature of a one-parameter group of transformations in phase space. Choose a state of the system, a point in phase space, as the starting point. There is a curve in phase space through this point on which the value of the dynamical variable, G, is a constant, but its canonically conjugate dynamical variable, $\bar{G}$, changes its value from point to point. We can use the change in the value of $\bar{G}$, call it $\varepsilon$, to label the points on the curve. The transformations generated by G take us from state to state along such a curve. We also can determine the change in any dynamical variable under the transformations generated by G.

We summarize the method for carrying out these calculations, citing references for details.[12, 13] Let $\varepsilon$ be the parameter, with $\varepsilon = 0$ corresponding to the starting point in phase space; $\varepsilon$ has physical dimensions of Action/G. Let F be any dynamical phase space variable; in particular, the analysis applies to canonical coordinates and momenta.

$$\frac{dF}{d\varepsilon} = \{F, G\}. \tag{11}$$

For example, we first let F be x and then $p_x$, and obtain

$$\frac{dx}{d\varepsilon} = \{x, G\} = \frac{\partial G}{\partial p_x}, \text{ and}$$
$$\frac{dp_x}{d\varepsilon} = \{p_x, G\} = -\frac{\partial G}{\partial x}, \text{respectively.} \tag{12}$$

Let us consider the special case where G is the angular momentum $\ell = xp_y - yp_x$. In this case, Eq. (12) becomes:

$$\frac{dx}{d\varepsilon} = \{x, L\} = \frac{\partial L}{\partial p_x} = -y, \text{ and}$$

$$\frac{dp_x}{d\varepsilon} = \{p_x, L\} = -\frac{\partial L}{\partial x} = -p_y, \text{ respectively.} \tag{12a}$$

Similarly, $\frac{dy}{d\varepsilon} = x$ and $\frac{dp_y}{d\varepsilon} = p_x$.

Note that, in this special case, the equations governing the transformation of the coordinates separate from the equations governing the transformation of the momenta, and moreover, the two sets of equations are formally identical. The solutions are

$$\begin{pmatrix} x(\varepsilon) \\ y(\varepsilon) \end{pmatrix} = \begin{bmatrix} \cos\varepsilon & -\sin\varepsilon \\ \sin\varepsilon & \cos\varepsilon \end{bmatrix} \begin{pmatrix} x(0) \\ y(0) \end{pmatrix}, \tag{12b}$$

and similarly for the momenta. The reader might recognize that Eq. (12b) cooresponds to ordinary rotations in the x-y plane, the rotation angle serves as the parameter as claimed earlier.

Similar equations can be found for other pairs of canonical variables. Upon integration with respect to ε, with the given initial values at ε = 0, we obtain the finite changes in the dynamical variables. We consider transformations generated by α. Numerical integration of the equations for the old coordinates and momenta is used to compute the transformation of phase space points. Because the phase space, even for this simple system is four-dimensional, it is better to use a reduced phase space such as the x-y plane or a three dimensional space whose axes are r, $p_x$, and $p_y$ for computer generated pictures of how a point in phase space moves under these transformations. Simpler than the effect of canonical transformations generated by α on the old coordinates and momenta, are their effect on the Hamiltonian and the angular momentum. Because the Poisson bracket of H with α vanishes, H is invariant under these transformations as we would expect. Also, {α, L} = 1 implies that L(ε) = L(0) + ε; so the transformations induce

"translations" in the conjugate angular momentum variable, again as we would expect. Physically, the energy is unchanged, but the angular momentum varies continuously between $-L_{max}$ and $+L_{max}$. In global terms, an elliptical orbit is transformed into another with the same major axis, but a different eccentricity, while preserving the direction of the major axis in space. The changes thus produced can be animated and are available for download.[14] Interested readers can think of several additional computer graphics projects along similar lines to help them consolidate these fundamental ideas of classical dynamics.

**II.    The Harmonic Oscillator**

For the isotropic two-dimensional harmonic oscillator, we also have elliptical orbits, and a given energy (proportional to the sum of the squares of the major and minor axes) corresponds to a range of angular momentum values, or alternatively, eccentricities. In this case, however, the center of force is at the geometric center of the ellipse. Hence, we do not expect an extra conserved *vector* to represent properly the dynamical symmetry of the oscillator. (Which way could such a vector point? Either of the two opposite directions along the major axis would be equivalent.) The object that best captures the degeneracy arising from the dynamical symmetry is a symmetric second rank tensor.[10] For convenience, let us choose units so that the mass, m = 1, and the spring constant, k = 1. We will denote the tensor (defined below) by A and write it in matrix form. Let the Cartesian coordinates be $x_i$ and the corresponding momentum components be $p_i$. It is not difficult to check that

$$A_{ij} = \frac{1}{2}(x_i x_j + p_i p_j) \tag{13}$$

are conserved in this system. For two degrees of freedom, there are three independent components: $A_{11}$, $A_{22}$, and $A_{12} = A_{s1}$. However, the trace of the tensor A is the Hamiltonian, and the determinant of the tensor is ($L^2/4$). Hence, as is to be expected, given the conservation of energy and angular momentum, there is only one more independent conserved quantity that the tensor A represents.

We digress for a moment to make a counting argument for central forces in n-dimensional space. The number of components of angular momentum ($L_{ij} = x_i p_j - x_j p_i$) in n-dimensions is n(n - 1)/2. The maximum number of independent (and time-independent) conserved functions in phase space is (2n - 1). Thus the search for dynamically conserved quantities besides the angular momentum components and energy is interesting only in two and three dimensions. Quite apart from the practical interest in two and three dimensions, physical considerations of the evolution of the system point in phase space also pick out these low dimensional problems as special. Of course, we do not mean to discount the purely theoretical value of investigating the more general cases. Moreover, as remarked earlier, because we understand the implications of the conservation of angular momentum in classical dynamics, there is no loss of generality in reducing the three-dimensional problem to the two dimensions of the plane of motion.

Returning now to the two-dimensional isotropic harmonic oscillator, we can again construct a single dynamical variable that represents the orientation of the elliptical orbit in the plane of motion, which we characterize as either the x-y plane or equivalently the $x_1$-$x_2$ plane. To see in detail how this construction can be done, note that each of the components of A is conserved, and hence it may be evaluated at any point in the orbit. Consider an elliptical orbit as shown in Fig. 1 whose major axis makes an angle β with the positive $x_1$ axis; the natural range of β may be taken to be $(-\frac{\pi}{2}, \frac{\pi}{2}]$. A little thought shows that at the point P in Fig. 1, the co-ordinate and momentum components are as follows:

$$x_1 = a \cos b \qquad x_2 = a \sin b \qquad (14a)$$

$$p_1 = -b \sin b, \qquad p_2 = b \cos b, \qquad (14b)$$

where a and b are respectively the semi-major and semi-minor axes of the ellipse; we have assumed that the orbit is traversed in the positive or counterclockwise sense. Hence,

$$A_{11} = \frac{1}{2}\left(a^2 \cos^2 \beta + b^2 \sin^2 \beta\right), \qquad 15a)$$

$$A_{12} = A_{21} = \frac{1}{2}\left(a^2 - b^2\right) \cos \beta \sin \beta, \qquad 15b)$$

and

$$A_{22} = \frac{1}{2}\left(a^2 \sin^2 \beta + b^2 \cos^2 \beta\right). \qquad 15c)$$

A little algebra to eliminate a and b yields

$$\beta(x_1, x_2, p_1, p_2) = \frac{1}{2} \tan^{-1}\left(\frac{2 A_{12}}{A_{11} - A_{22}}\right). \qquad (16)$$

Of course, β is a constant of the motion, because it is expressed as a function only of the other constants of the motion. As such it has a vanishing Poisson bracket with the Hamiltonian. A straightforward calculation shows that $\{\beta, L\} = 1$. As in Sec. I, the effect of the transformation generated by β is to produce local translations in L while leaving the energy invariant. In configuration space, these transformations take one elliptical orbit to another with the same orientation, the same value of the sum of the squares of the major and minor axes, but with a different ratio between the axes. All the other developments are analogous to the case of the inverse-square-force problem. The reader can play with computer graphics to better understand the effect of the one-parameter group of transformations on the dynamical variables in this familiar context as well.

## ACKNOWLEDGMENTS

We thank E. J. Saletan and J. V. José for helpful correspondence. We also thank an anonymous referee who read the manuscript with care, and made a number of suggestions for greater clarity or elaboration.

References

1. H. Goldstein, "Prehistory of the 'Runge-Lenz' vector," Am. J. Phys. **43**, 737-738 (1975); ibid., "More prehistory of the Laplace or Runge-Lenz vector," Am. J. Phys. **44**, 1123-1124 (1976). The references in these two papers, while not exhaustive, point to some very good early work.

2. L. Basano and A. Bianchi, "Rutherford's scattering formula via the Runge-Lenz vector," Am. J. Phys. **48**, 400- 401 (1980).

3. R. P. Patera, "Momentum-space derivation of Runge-Lenz vector," Am. J. Phys. **49**, 593-594 (1981).

4. H. Kaplan, "The Runge-Lenz vector as an 'extra' constant of motion," Am. J. Phys. **54**, 157-161 (1986).

5. A. C. Chen, "Coulomb-Kepler problem and the harmonic oscillator," Am. J. Phys. **55**, 250-252 (1987).

6. X. L. Yang, M. Lieber, and F. T. Chan, "The Runge-Lenz vector for the two-dimensional hydrogen atom," Am. J. Phys. **59**, 231-232 (1991).

7. C. E. Aguiar and M. F. Barroso, "The Runge-Lenz vector and the perturbed Rutherford scattering," Am. J. Phys. **64**, 1042-1048 (1996).

8. D. R. Brill and D. Goel, "Light bending and perihelion precession: A unified approach," Am. J. Phys. **67**, 316-319 (1999).

9. H. Bacry, H. Ruegg, and J.-M. Souriau, "Dynamical groups and spherical potentials in classical mechanics," Commun. Math. Phys. **3**, 323-333 (1966).

10. H. V. McIntosh, "Symmetry and degeneracy" in *Group Theory and Its Applications,* edited by E. M. Loebl (Academic Press, New York, 1971), Vol. II, pp. 75-144.

11. N. Mukunda, "Realizations of Lie algebras in classical mechanics," J. Math. Phys. **8**, 1069-1072 (1967); the present paper is closest in spirit to this reference.

12. E. C. G. Sudarshan and N. Mukunda, *Classical Dynamics: A Modern Perspective* (John Wiley, NY, 1974), pp. 30-77.

13. J. V. José and E. J. Salaten, *Classical Dynamics: A Contemporary Approach* (Cambridge University Press, 1998), pp. 248-274.

14. Animated plots can be seen at <http://www.amherst.edu/~physics/DynamicalSymmetries/>. These plots were generated by using standard packages, for example Mathematica.

Figure Caption

Fig. 1. The elliptical orbit of a two-dimensional harmonic oscillator with the major axis inclined at an angle β to the $x_1$ axis. For convenience, we use units in which the mass and spring constant are set to unity.

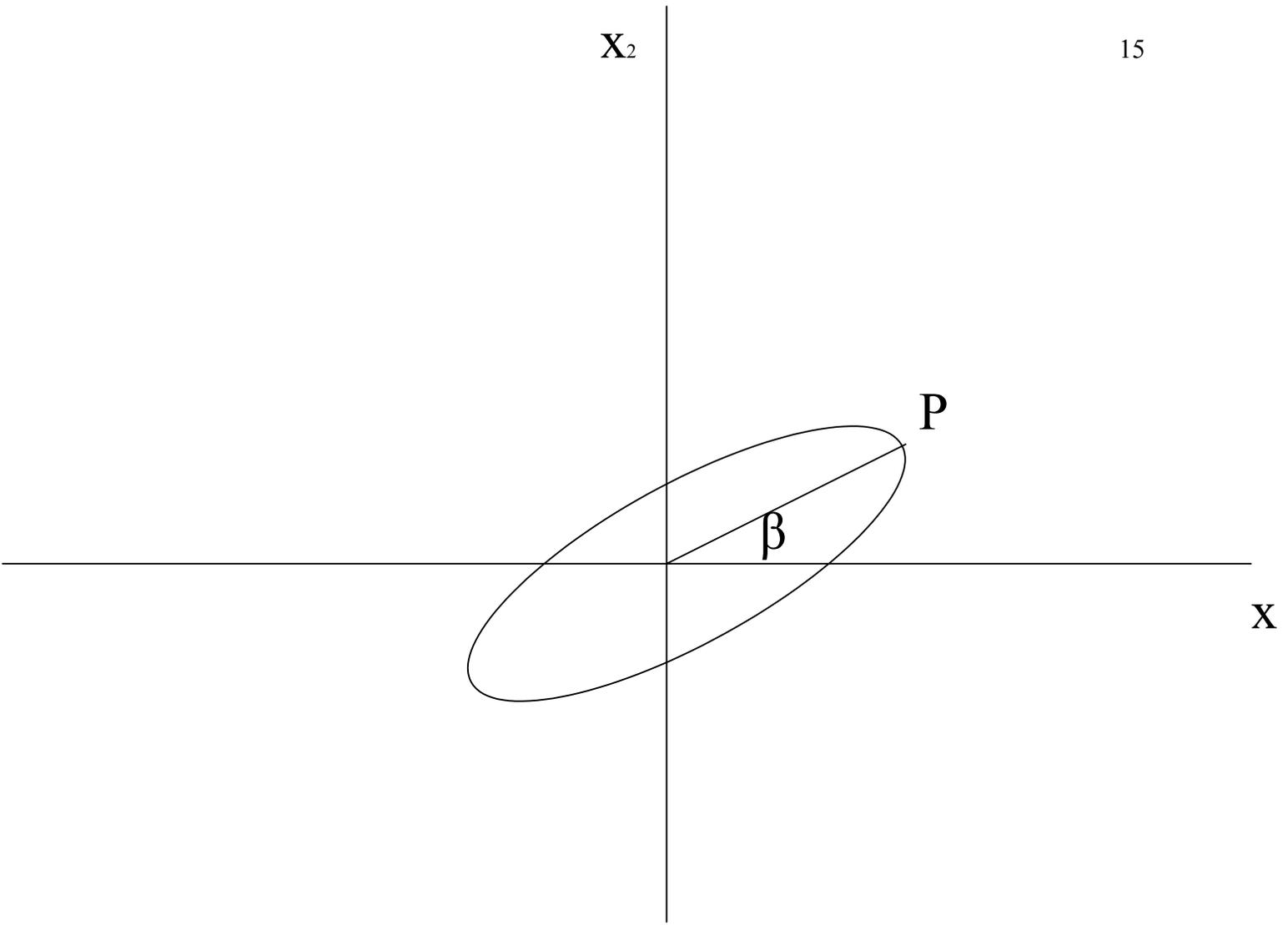